\documentstyle [12pt,a4] {article}
\begin{document}
\vskip .5in
\begin{center}
{\Large \bf Beyond Standard Model : Report of Working Group II}\\[25pt]
{\large\sl Workshop on High Energy Particle Physics 3 (Madras, Jan.
10--23, 1994)} \\[50pt]
{\bf Anjan S. Joshipura}\\ \vskip .2in
 Theory Group, Physical Research Laboratory\\Navrangpura,
Ahmedabad 380009, India\\[15pt]
and\\[15pt]
{\bf Probir Roy}\\ \vskip .2in
 Theory Group, Tata Inst. of Fundamental Research\\Homi Bhaba Road,
Bombay 400005, India\\[40pt]
Participants:\\
{\it K.S. Babu, B. Brahmachari, C. Burgess, G. Datta, S.
Goswami, A. Joshipura\\ A. Kundu, Mohan Narayan, S. Nayak, M. V. N. Murthy,
M.K. Parida\\ G. Rajasekaran, S. Rindani, Probir Roy, J.W.F. Valle}
\end{center}
\begin{abstract}
Working group II at WHEPP3  concentrated
on issues related to the supersymmetric standard model as well as SUSY
GUTS and neutrino properties. The projects identified by various
working groups as well as progress made in them since WHEPP3 are
briefly reviewed.
\end{abstract}
\newpage
Working group II (WGII) identified definite topics each of which was
intensively discussed within the corresponding subgroup during the
workshop. Significant progress was made in some of them and some of
the projects have been completed in the meantime. The following is the
list of the projects addressed by WGII:
\begin{itemize}
\item Evolution of $R$ parity violating couplings ({\it
B. Brahmachari and P. Roy})
\item Beyond $S$, $T$ and $U$ ({\it A. Kundu and P. Roy})
\item Neutrino masses and proton lifetime in SUSY $SO(10)$ ({\it
K.S. Babu, M.K. Parida and G. Rajasekaran})
\item Degenerate neutrinos({\it K.S. Babu, C. Burgess, A.S.
Joshipura, S. Rindani, J.W.F. Valle})
\item Solar and atmospheric neutrino problems with three generations ({\it G.
Datta, S. Goswami, A. Joshipura, M.V.N. Murthy, Mohan Narayan, G. Rajasekaran
and S. Rindani })
\item Magnetic moments for heavy neutrinos ({\it K.S. Babu, S.
N. Nayak and P. Roy })
\item Extraction of neutrino magnetic moment from experiments
  ({\it M.V.N. Murthy, G. Rajasekaran and S. Rindani})
\item Evolution of couplings in SUSY LR model ({\it B. Brahmachari})
\end{itemize}
\section{Evolution of $R$ violating couplings}
Brahmachari and Roy \cite{br} studied the evolution of the baryon number
and $R$-parity violating Yukawa couplings in the supersymmetric
standard model and derived bounds on them from the
requirement of perturbative unitarity. They added the following
terms to the superpotential of the minimal supersymmetric
standard model (MSSM):
\begin{equation}
{\cal L}=\lambda^{'''}_{ijk}(D^c_iD^c_j U^c_k),
\end{equation}
where $U^c,D^c$ denote the anti quark superfields and $i,j,k$ are
generation indices. These terms violate both $R$ parity and the baryon
number.  Unlike the analogous lepton number violating terms, the
presence of the above terms by themselves is not significantly
constrained from low energy considerations. Interesting bounds on
these couplings can nevertheless be obtained by requiring that all the
Yukawa couplings $Y$ remain less than unity till the grand unification
scale $M_U\sim 2 \times 10^{16}$ GeV is reached. Assuming only
$\lambda^{'''}_{133}$ and $\lambda^{'''}_{233}$ to be large, they set
up the RG equations for the relevant couplings.  The requirement of
perturbative unitarity was shown to lead to an upper bound in the
range 0.5-0.6 on the baryon number violating Yukawa couplings, the
exact value being dependent on the top quark mass as well as on the
ratio $\tan\beta$ of the Higgs vevs. It was also shown that the fixed
point value of the top Yukawa coupling was somewhat reduced compared
to that in the MSSM because of the presence of the additional baryon
number violating Yukawa couplings.

\section{Oblique corrections beyond the linear approximation}

The lectures of C. Burgess (included in the proceedings) discussed the
question of going beyond the linear $Q^2$-expansion approximation used
in 1-loop oblique electroweak radiative corrections and the new
oblique parameters $V, W, X$ and $Y$.  A. Kundu and P. Roy afterwards
examined the same question in another way.  They formulated
$q^2$-expansion independent definitions of $S,T,U$ which are different
from those of Burgess, Maksymic and London.  The difference concerns
the broken and custodial symmetry contents of these parameters.  Kundu
and Roy extended the Peskin-Takeuchi definitions beyond the linear
approximation whereas BML did the same with the Marciano-Rosner
definitions.  The BML and KR definitions differ in terms of weak
isospin and hypercharge breaking properties; the choice of definition
can be regarded as a matter of convenience but different definitions
mean different physical quantities.  Kundu and Roy have further found
the organizing principle behind the $q^2$-expansion approximation ---
namely that it is needed in calculating the $Z$- and $W$-wavefunction
renormalization constants.  Stringent experimental bounds are obtained
on $S,T$ and $U$ without reference to this approximate procedure.  The
new oblique parameters $V,W,X,Y$ have been bounded [2] experimentally
within quite tight ranges for the first three.

\section{Neutrino masses in SUSY $SO(10)$}
K. S. Babu, M. K. Parida and G. Rajasekaran looked at the issue of
obtaining neutrino masses in the experimentally interesting range in
the context of supersymmetric $SO(10)$ models. The neutrino masses
needed for solving the solar neutrino problem arise naturally in and
$SO(10)$ if the Majorana masses of the right handed neutrinos
are in the intermediate range $\sim 10^{10}$ GeV \cite{lang}. The
generation of such masses through the vacuum expectation values of the
chargeless scalars in the $126+\bar{126}$ representation in SUSY
$SO(10)$ models requires \cite{desh} some assumptions of the extended
survival hypothesis. The aim of the project was to provide an
alternative mechanism for generating the right handed neutrino masses
in the intermediate energy range.

The following breaking chain was considered:
\begin{equation}
 SO(10)\longrightarrow G_I\longrightarrow G_{SM},
\end{equation}
where $G_I=SU(3)_c\times SU(2)_L\times SU(2)_R\times U(1)_{B-L}$ or
$SU(4)_c\times SU(2)_L\times SU(2)_R$ and was assumed to break at a scale
$M_I\sim 10^{14} GeV$. Although the representation
$126+\bar{126}$ was present it did not acquire a vev. The right
handed neutrino masses were induced by the presence of the
$16+\bar{16}$ representation to be
\begin{equation}
M_{N_R}\sim \frac{M_I^2}{M_U}.
\end{equation}
This could be significantly lower than the value $M_{N_R}\sim
M_U\sim 10^{16}$ GeV permitted in a single step breaking.
\section{Degenerate neutrinos}
It has recently been realized \cite{moh,asj1} that simultaneous
solutions of the solar and atmospheric neutrino deficits as well as of
the dark matter problem with a hot component of about 30\% require almost
degenerate masses for the three neutrinos.  Such a spectrum was shown
to arise in a natural manner in left right symmetric models augmented
with a suitable generation symmetry \cite{moh,asj1}. The aim of the
working group was to discuss issues related to the construction of
realistic grand unified models following the scenario proposed in
refs. \cite{moh,asj1}. In particular one should obtain (a) the common
degenerate mass in the eV range (b) mass splittings appropriate for
the solar and the atmospheric neutrino problems and (c) the right
mixing pattern.  The required mass splitting arise naturally
\cite{asj1} if the Dirac masses for the neutrinos coincide with the up
quark masses as in the simplest $SO(10)$. In this case, one obtains
\begin{equation}
\frac{|\Delta_{21}|}{|\Delta_{32}|}\approx
\left(\frac{m_c}{m_t}\right)^2\approx (1-3)\times  10^{-4}.
\end{equation}
This nicely reproduces the hierarchy required to simultaneously
solve the solar and atmospheric neutrino problems. The problem to
be addressed was to obtain this prediction in a complete model
based on $SO(10)$ preserving other successful features.

While a complete model is still lacking, significant progress was made
by the members of the working group \cite{cliff,ion,asj2} as well as
others \cite{moh2} in the construction of realistic models. In
particular, Valle and Ioannissyan constructed a model based on
$SO(10)$ with a horizontal $SU(2)$ symmetry. In their model, the up
quark mass matrix coincided with the Dirac neutrino masses leading to
eq.(4).  The down quark mass matrix is however not proportional to the
charged lepton masses. This allows enough freedom to obtain the
required mixing pattern.  A similar model was also proposed by Caldwell
and Mohapatra \cite{moh2}.  Bamert and Burgess worked out a scenario
which contained a singlet fermion in addition to the three left and
right handed neutrinos.  A horizontal SU(2) symmetry was introduced to
obtain the degenerate spectrum.  The couplings involving the singlet
fermion break the horizontal symmetry and lead to a departure from the
degeneracy in neutrino masses.  The singlet fermion was moreover used in
the context of the left right symmetric theory \cite{asj2} in order to
understand the difference between the quark and leptonic mixing angles
in scenarios with almost degenerate neutrinos. The singlet also played
a crucial role in generating the required mass pattern among
neutrinos in this scenario.

\section{Solar and atmospheric neutrino problems with three generations}

The understanding of the solar and atmospheric neutrino deficits in
terms of neutrino oscillations seems to require two vastly
different values for the (mass)$^2$ difference among  neutrinos.
Thus at least two neutrinos need to be massive and analysis of the
solar and atmospheric neutrino data in terms of three generations becomes
interesting. Such an analysis was carried out earlier \cite{krast,fogli}
assuming the MSW mechanism to be responsible for the solar
neutrino conversion. This working group looked at a complimentary
scenario in which two of the neutrinos were assumed to be almost
degenerate with very small (mass)$^2$ difference $\sim 10^{-10}$
(eV)$^2$ while the other (mass)$^2$ difference was assumed to
be in the range  $\sim 10^{-2}-10^{-3}$
(eV)$^2$. Thus the vacuum oscillations are responsible for both
the solar and the atmospheric neutrino deficit. Since two of the
relevant (mass)$^2$ differences show hierarchy, the oscillation
probabilities involve only one more mixing angle compared to the
case of two generations \cite{krast,fogli}. Fixing this mixing
angle ($\theta_{\mu\tau}$) to be in the range appropriate for the atmospheric
neutrino problem, restrictions on other mixing angle (namely
$\theta_{e\mu}$ ) and the (mass)$^2$ difference $\Delta_{e\mu}$ were
determined from the data on solar neutrino deficit.

\section{Neutrino magnetic moment}

Two different problems were analyzed in connection with the neutrino
magnetic moment. One was the issue of a large magnetic moment of a
very heavy neutrino. Since the magnetic moment of fermion turns out to
be proportional to its mass in a number of situations, it is
interesting to ask if the magnetic moments of heavy singlet neutrinos
can be large enough to dominate over their point couplings to $W$ and
$Z$ induced by mixing with the light neutrinos. The typical magnetic
moment of a very heavy right handed neutrino $N$ was estimated from
the one-loop graph and the mass-dependence was seen to come through
the factor $m_LM_N (M^2_N+M^2_W)^{-1}$, where the $W$ couples to $\ell$
and $N$, so that there was no enhancement for $M_N \gg M_W$.  Thus it
was found that, contrary to naive expectation, the point couplings
always dominated over the magnetic moment couplings.

The conventional procedure of extracting information on the neutrino
magnetic moment coupling from the data on $\nu\;e$ scattering was
questioned. In order to extract the magnetic moment from the data, one
conventionally writes an effective phenomenological term
$\kappa_{\nu}\sigma_{\mu\lambda}q_{\lambda}/m$ in the calculation of
the neutrino electron scattering. An analogous treatment of the $e-p$
scattering has been shown to lead to a drastic overestimation of the
QED radiative corrections \cite{rahul}. By the same token, the
inclusion of the neutrino magnetic moment term through the Pauli term
must lead to wrong results at some energy scale.  The main issue was to
determine the relevant scale where the Pauli approximation breaks
down. The suggestion was to do a detailed calculation of $\nu e$
scattering in specific model which leads to large magnetic moment and
compare it with the phenomenological result obtained assuming the
Pauli term as is conventionally done.

\section{Evolution of couplings in SUSY LR model}

B. Brahmachari studied the 1-loop evolution of Yukawa copulings in the
minimal supersymmetric left-right model.  He found \cite{biswa} a
fixed point behaviour in the top Yukawa coupling that was rather
analogous to the one one in the MSSM.  He was able to explicitly
exhibit the dependence of the fixed point solution of $Y_t(m_t)$ on
the right-symmetry breaking scale.  The predicted top mass value in
this scheme was between 168 and 174 GeV.  Brahmachari was also able to
fix the value of the Majorana Yukawa coupling which is otherwise a
free parameter.

\end{document}